\documentclass[generic,preprint]{imsart}
\usepackage{natbib}
\usepackage{graphicx}
\usepackage{array}
\usepackage{multirow}
\usepackage{amsthm}
\usepackage{amsmath}
\usepackage{amsfonts}
\usepackage{amssymb}
\usepackage{pgf}
\usepackage{bm}
\usepackage{paralist}
\usepackage{pdflscape}
\usepackage{rotating}
\usepackage{scalefnt}
\usepackage{comment}
\usepackage{booktabs,caption}
\usepackage[flushleft]{threeparttable}
\usepackage{booktabs,siunitx}
\usepackage{graphicx}
\usepackage{chngpage}
\usepackage[skip=20pt]{caption}

\usepackage{tikz}
\usepackage{ifthen}
\usetikzlibrary{arrows,shapes}
\usepackage{xr}

\startlocaldefs
\numberwithin{equation}{section}
\theoremstyle{plain}

\newcommand{\beq}{\begin{equation}}
\newcommand{\eeq}{\end{equation}}
\newcommand{\bea}{\begin{eqnarray}}
\newcommand{\eea}{\end{eqnarray}}
\newcommand{\bit}{\begin{itemize}}
\newcommand{\eit}{\end{itemize}}
\newcommand{\ben}{\begin{enumerate}}
\newcommand{\een}{\end{enumerate}}
\newcommand{\bpm}{\begin{pmatrix}}
\newcommand{\epm}{\end{pmatrix}}
\newcommand{\bbm}{\begin{bmatrix}}
\newcommand{\ebm}{\end{bmatrix}}

\endlocaldefs
\RequirePackage[OT1]{fontenc}
\RequirePackage{amsthm,amsmath}
\RequirePackage[colorlinks, citecolor=blue, urlcolor=blue]{hyperref}

\begin{document}

\begin{frontmatter}
\title{Bayesian estimation of large dimensional time varying VARs using copulas \thanks{The first author acknowledges partial financial support from the Bank of Greece. We
are grateful to Dr Xinghai Yao for excellent research assistance.}}
\runtitle{Large Bayesian TVP-VARs}

\begin{aug}
\author{\fnms{Mike G.} \snm{Tsionas}$^{\dag}$ \ead[label=e1]{m.tsionas@lancaster.ac.uk}}\hskip .2cm
\author{\fnms{Marwan} \snm{Izzeldin}$^{\dag*}$ \ead[label=e2]{m.izzeldin@lancaster.ac.uk}}\hskip .2cm
\author{\fnms{Lorenzo} \snm{Trapani}$^{*}$\ead[label=e3]{lorenzo.trapani@nottingham.ac.uk}}
\runauthor{M. Tsionas, M. Izzeldin and L. Trapani}

\affiliation{Lancaster University \thanksmark{m1} and University of Nottingham\thanksmark{m2}}

\address{$^{\dag}$Lancaster University Management School\\
\printead{e1}\\
}

\address{$^{\dag*}$Lancaster University Management School\\
\printead{e2}\\
}

\address{$^{*}$University of Nottingham\\
\printead{e3}\\
}
\end{aug}

\begin{abstract}

This paper provides a simple, yet reliable, alternative to the (Bayesian) estimation of large multivariate VARs with time variation in the conditional mean equations and/or in the covariance structure. With our new methodology, the original multivariate, $n$-dimensional model is
treated as a set of $n$ univariate estimation problems, and cross-dependence is handled through the use of a copula. Thus, only univariate distribution functions are needed when estimating the individual equations, which are often available
in closed form, and easy to handle with MCMC (or other techniques). Estimation is carried out in parallel for the individual equations. Thereafter, the individual posteriors are combined with the copula, so obtaining a joint posterior which can be easily resampled. We illustrate our approach by applying it to a large time-varying parameter VAR with $25$ macroeconomic variables.
\bigskip

\end{abstract}


\end{frontmatter}


\section{Introduction\label{introduction}}

Following the seminal contributions by \citet{sims80} and %
\citet{litterman1986}, Vector AutoRegression (VAR) models and their variants are now ubiquitously applied to multivariate time
series, as a flexible alternative to structural models. There is now a
huge body of literature on both theory and applications: useful
surveys are provided \textit{inter alia} by \citet{watson94} and %
\citet{lutkepohl91}. \newline
Although in their
standard form VARs already offer a relatively flexible modelling approach,
extensions have been considered to accommodate time variation. This may
occur in the coefficients of the conditional mean equations (see e.g. %
\citealp{doan}; \citealp{canova}; \citealp{sims93}; \citealp{stock96}; and %
\citealp{cogley}), so affording a flexible alternative to models with abrupt
breaks, known as the Time-Varying Parameter VAR (TVP-VAR). Time variation
has also been considered in the covariance matrix of the error term, thereby
allowing for time-varying heteroskedasticity.
Following seminal papers by \citet{uhlig}, \citet{cogley} and \citet{primiceri}, recent examples where the assumption of
homoskedasticity has been relaxed include \citet{koop13} and %
\citet{koop19}, who attempt to reduce the dimensionality issue essentially by
imposing a factor structure onto the volatilities - see also %
\citet{clark2011}, \citet{carriero2015}%
, \citet{clark2015} and \citet{carriero2012}. In a recent landmark paper, \citet{carriero2016large} propose a far less restrictive set-up, which
allows for fully Bayesian inference without imposing restrictions on the
form of the heteroskedasticity. 

\bigskip

Across the extensive literature on multivariate models, virtually all studies have one
issue in common: the dimensionality of the model and the computational burden it brings about. On the one hand, unless the
number of variables involved in the model is relatively large, omitted
variable bias may impair the forecasting ability of the model (see %
\citealp{giannone2006}). \citet{carriero2016large} make a compelling case
for the superior performance of large dimensional VARs. On the other hand, computational
difficulties may arise when there are a large number of variables and, more substantially, over-parameterisation can occur. Thus, the
literature has focused on finding techniques that allow for the estimation
of large VARs: see \citet{banbura} for an excellent review of the various approaches which have been proposed. 
In the case of homoskedastic VARs, the dimensionality issue can be handled by the choice of appropriate (conjugate) prior distributions, as shown by \citet{banbura} who successfully apply their technique to the estimation of a VAR with 130 variables. Conversely, in the case of a heteroskedastic VAR, this is no longer possible, and the computational burden cannot be resolved through the choice of an appropriate prior. As explained in \citet{carriero2016large}, heteroskedasticity invalidates the so-called \textquotedblleft symmetry\textquotedblright across equations that characterises homoskedastic VARs. Such property entails that a homoskedastic VAR is a SUR model where the regressors are the same across all equations; in turn, this entails that the covariance matrix of the VAR coefficients has a Kronecker structure, which makes estimation much simpler than if one had to deal with large matrices that do not have a simplifying structure. Few contributions consider estimation of a large VAR with time variation in both the coefficients of the conditional mean equations and of the covariance matrix of the error term. \citet{koop13} and \citet{koop19} propose an estimation technique for large, possibly heteroskedastic TVP-VARs, which essentially relies on the Kalman filter. However, their approach is not fully Bayesian, and it is liable to mis-specification issues if the assumed model for
coefficient variation is not correct, also being, in practice, limited
in dealing with the dimensionality issues - we refer however to a recent
contribution by \citet{venditti} which, through a non-parametric approach,
solves these issues. \citet{carriero2016large} solve the problem of fully Bayesian estimation of large VARs with heterskedasticity, by proposing a new estimation algorithm which is shown to perform very well in out-of-sample forecasting and also in structural analysis. However, their paper does not consider the presence of time varying coefficients in the VAR specification.

\bigskip

\textit{Proposed methodology and main contribution of this paper}

\bigskip

This paper proposes a copula-based Bayesian estimation methodology for large TVP-VARs with heteroskedasticity. Similarly to \citet{carriero2016large}, our estimators are fully Bayesian, thus allowing for the computation of the uncertainty around all estimators. \newline
Full details of our approach are in Section \ref{methodology}; here, we give
a heuristic preview of our methodology. Given a multivariate model of
(possibly) very large dimension $n$, we reduce it into $n$ univariate
models, which are more easily handled. In order to recover the cross-dependence
among equations, we use a copula-like term. In consequence, the likelihood
function in our system factors into the likelihoods of the individual
autoregressive models, plus the likelihood of the copula term. Thereby we are able to obtain a posterior where each set of parameters (the $n$
sets corresponding to the $n$ equations, and the set corresponding to the
copula) is, conditional on the sample, independently distributed of the
other sets. Therefore, from a computational point of view, each univariate
problem is dealt with separately, which greatly reduces the computational burden. In this respect, our idea of breaking down the multivariate estimation problem into separate univariate problems is similar to the approach for fixed parameter, homoskedastic VARs developed in \citet{korobilis2019adaptive}, although in our case we allow for time variation in the conditional mean and variance. The use of copulas to model dependence has also been considered by the literature in a Bayesian context (see e.g. \citealp{gruber2015} and \citealp{gruber2018}), including non-parametric Bayesian analysis (we refer, \textit{inter alia}, to the contributions by  %
\citet{rodriguez2008bayesian}, \citet{taddy2010autoregressive}, %
\citet{nieto2012time}, \citet{di2013simple}, \citet{bassetti2014beta} and %
\citet{nieto2016bayesian}). \\
Our approach allows for great flexibility in the specification of the univariate models. For example,
in the simplest version of our methodology, each series is modelled as an
AR(1) model. However, given that more sophisticated model selection tools
may be desirable to construct the univariate models, we develop an approach based
on a model selection technique known as Bayesian compression (see %
\citealp{dunson}). Moreover, whilst the focus of this paper in on TVP-VARs with heteroskedasticity, our approach can be also used to estimate other multivariate models such as e.g. VARMAs and Multivariate Stochastic Volatility models (MSV). In particular, in appendix we carry out an empirical exercise using VARMAs, to illustrate the computational convenience of our approach. Also, in another contribution (\citealp{itt2019}), we apply our methodology to large MSV models for financial variables. Here, we illustrate our methodology by estimating a large TVP-VAR model with
possible heteroskedasticity, using the same data as \citet{koop13}. 

\bigskip

The remainder of the paper is organised as follows. Our methodology is spelt
out in Section \ref{methodology}. The empirical application is in Section \ref%
{empirics}; we also report a further application to VARMAs in Appendix \ref{varma}. Section \ref{conclusions} concludes. 

\section{Methodology\label{methodology}}
We begin by introducing the main model and some notation. We consider the
TVP-VAR($p$) 
\begin{equation}
\mathbf{y}_{t}=\sum_{j=1}^{p}A_{t,j}\mathbf{y}_{t-j}+\mathbf{u}_{t},\text{ \
\ }p+1\leq t\leq T,  \label{var}
\end{equation}%
where $\mathbf{y}_{t}$ is an $n\times 1$ vector and $\mathbf{u}_{t}$ is a
zero mean, Gaussian process with possibly time varying variance - we discuss
the specification of the second moment later on. Model (%
\ref{var}) could be extended to consider e.g. exogenous regressors, latent
regressors such as common factors, or deterministics such as a
constant, (linear or nonlinear) trends and seasonal dummies. Also, (\ref{var}%
) could also have an MA($q$) structure, in the spirit of \citet{chan2013};
or it could have no autoregressive structure at all, and only time varying
heteroskedasticity as in the case of \citet{creal}. We prefer to focus on a
simpler specification, so that the discussion is not overshadowed by the
algebra. Similarly, the assumption that $\mathbf{u}_{t}$ is Gaussian is made
only for simplicity. Note that, even in this simple set-up, the number of
parameters grows rapidly with $p$ and $n$, whence the dimensionality issue.

\subsection{Theory: the univariate equations, the copula and the likelihood
function \label{copula}}


\textit{The univariate equations}

\bigskip 

In the context of (\ref{var}), we consider the following univariate AR($p$)
models%
\begin{equation}
y_{i,t}=\sum_{j=1}^{p}\beta _{i,t,j}y_{i,t-j}+u_{i,t}=\beta _{i,t}^{\left(
1\right) \prime }z_{i,t}^{\left( 1\right) }+u_{i,t},  \label{univ}
\end{equation}%
for $1\leq i\leq n$ with $u_{i,t}=e^{h_{i,t}/2}u_{i,t}^{*}$, where $%
u_{i,t}^{\ast }\sim i.i.d.N(0,1)$ and 
\begin{equation}
h_{i,t}=\alpha_{i}+\gamma_{i}h_{i,t-1}+e_{i,t},  \label{garch}
\end{equation}%
with $e_{i,t}\sim i.i.d.N(0,\delta _{i})$. As noted above, (\ref{univ})
can be extended and/or modified to incorporate e.g. a different number of
lags $p_{i}$ for each unit, an MA($q_{i}$) component, exogenous regressors,
deterministics, (conditional or unconditional) heteroskedasticity, etc..
Similarly, (\ref{garch}) could be replaced e.g. by a GARCH specification to
allow for conditional heteroskedasticity (see also the discussion in %
\citealp{uhlig} on the relative merits of possible specifications for time
heteroskedasticity).

Given that in (\ref{univ}) $y_{i,t}$ is predicted using only its own past,
this may lead to a loss of predictive accuracy. A possible alternative would
be to use the Bayesian compression algorithm developed in \citet{dunson}. In
particular, we consider the specification%
\begin{equation}
y_{i,t}=\beta _{i,t}^{\left( 2\right) \prime }z_{i,t}^{\left( 2\right)
}+u_{i,t},  \label{tvp-univ-2}
\end{equation}%
with $u_{i,t}$ still satisfying (\ref{garch}). As in (\ref{univ}), $%
z_{i,t}^{\left( 2\right) }$ is a subset of the regressors in each equation
of the unrestricted VAR (say $\widetilde{z}_{i,t}$). However, in the case of
(\ref{tvp-univ-2}), the vector $z_{i,t}^{\left( 2\right) }$ can include lags of $%
y_{i,t}$ and also lags of $y_{j,t}$ for $j\neq i$. In order to select the
components of $z_{i,t}^{\left( 2\right) }$, \citet{dunson} suggest the
following technique. Let $z_{i,t}^{\left( 2\right) }=\Phi \widetilde{z}_{i,t}
$ with $\Phi $ a $p\times np$ matrix whose entries are defined as 
\begin{equation*}
\Phi _{ij}=\left\{ 
\begin{array}{c}
-\phi ^{-1/2} \\ 
0 \\ 
\phi ^{-1/2}%
\end{array}%
\right. \text{with probability }%
\begin{array}{c}
\phi ^{2} \\ 
2\phi \left( 1-\phi \right)  \\ 
\left( 1-\phi \right) ^{2}%
\end{array}%
,
\end{equation*}%
and $\phi $ and $p$ are drawn uniformly from $\left( 0.1,1\right) $ and $%
\left\{ 1,...,p^{\max }\right\} $, with $p^{\max }$ chosen so that the
marginal likelihood has a global peak. The matrix $\Phi $ is then normalised
via the Gram-Schmidt orthonormalisation - see \citet{dunson} for details.

\bigskip 

\textit{The copula}

\bigskip 

We now introduce the copula term to model dependence among the univariate
equations. Letting $X$ denote a continuous $k$-dimensional random variable
whose density is given by $f\left( x\right) $, it holds that%
\begin{equation}
f\left( x\right) =\left( \prod\limits_{j=1}^{k}f_{j}\left( x_{j}\right)
\right) c^{\ast }\left( v_{1},...,v_{k}\right) ,  \label{sklar}
\end{equation}%
where $f_{j}\left( x_{j}\right) $ is the density of the $j$-th coordinate of 
$X$, $v_{j}=F_{j}\left( x_{j}\right) =\int_{-\infty }^{x_{j}}f_{j}\left(
u\right) du$, and $c^{\ast }\left( v_{1},...,v_{k}\right) $ is the copula
density (which is unique since $X$ is continuous). This result is known as
Sklar's theorem (see \citealp{sklar1} and \citealp{sklar2}; see also the
book by \citealp{nelsen} for a comprehensive introduction to copulas).
Equation (\ref{sklar}) can equivalently be written as 
\begin{equation}
\ln f\left( x\right) =\sum_{j=1}^{k}\ln f_{j}\left( x_{j}\right) +\ln
c^{\ast }\left( v_{1},...,v_{k}\right) .  \label{log-sklar}
\end{equation}

\bigskip 

\textit{The likelihood function}

\bigskip 

We now turn to specifying the likelihood. Henceforth, we use $z_{i,t}$ as
short-hand for both $z_{i,t}^{\left( 1\right) }$ and $z_{i,t}^{\left(
2\right) }$; $\beta _{i,t}$ for both $\beta _{i,t}^{\left( 1\right) }$ and $%
\beta _{i,t}^{\left( 2\right) }$ in (\ref{univ}) and (\ref{tvp-univ-2})
respectively. We assume the following law of motion%
\begin{equation}
\beta _{i,t}=A_{\beta ,i}\beta _{i,t-1}+\epsilon _{i,t},  \label{motion}
\end{equation}%
with $\epsilon _{i,t}\sim i.i.d.N\left( 0,\Sigma _{i}\right) $, independent
across $i$. We point out that, in (\ref{motion}), we do not
impose the typical random walk model for the time-varying parameters (see
e.g. \citealp{koop13}), which makes our set-up more general. For simplicity,
we do not allow for time variation in any other parameter (i.e., we do not
allow for the copula parameters, or the coefficients in (\ref{garch}), to
vary over time). 

Let $b_{i}=\left( \alpha_{i},\gamma _{i},\delta _{i}\right) $. Then, the marginal
density of $y_{i,t}$ conditional on $z_{i,t}$ can be denoted as $f_{i}\left(
y_{i,t}|z_{i,t};\beta _{i,t},b_{i}\right) $ (we omit dependence on $A_{\beta
,i}$, $\Sigma _{i}$ and $\beta _{i,0}$ for short). Then, by (\ref{sklar}),
it holds that%
\begin{equation}
f_{i}\left( \mathbf{y}_{t}|z_{t};\beta _{i,t},b_{i}\right) =\left(
\prod\limits_{i=1}^{n}f_{i}\left( y_{i,t}|z_{i,t};\beta _{i,t},b_{i}\right)
\right) c^{\ast }\left( v_{1,t},...,v_{n,t}\right) ,
\label{conditional-joint}
\end{equation}%
having defined $v_{i,t}=\int_{-\infty }^{y_{i,t}}f_{i}\left( u|z_{i,t};\beta
_{i,t},b_{i}\right) du$, with $f_{i}\left( u |z_{i,t};\beta
_{i,t},b_{i}\right) $ denoting the density of $y_{i,t}$ conditional on $%
z_{i,t}$.

Although Sklar's theorem ensures that the copula density $c^{\ast }\left(
v_{1,t},...,v_{n,t}\right) $ is unique, it does not provide an expression
for it. One possibility would be to consider a non-parametric copula, and we
refer to \citet{scaillet2002}, \citet{ibragimov} and %
\citet{chen2007nonparametric}, and the references therein, for the relevant
theory in a time series context. In this paper, we choose a different
set-up. In particular, we consider a (parametric) Gaussian mixture copula
model (GMCM henceforth; see \citealp{tewari}) model, viz.%
\begin{equation}
c^{\ast }\left( v_{1,t},...,v_{n,t}\right) =\frac{\sum_{g=1}^{G}p_{g}f_{N}\left( 
\mathbf{y}_{t}|\mu _{g},\Omega _{g}\right)}{\prod_{g=1}^{G}p_{g}f_{N}\left( 
\mathbf{y}_{t}|\mu _{g},\Omega _{g}\right)} \equiv c\left( \mathbf{y}%
_{t}|\alpha \right) ,  \label{mixture-normal}
\end{equation}%
where: $\left\{ p_{g}\right\} _{g=1}^{G}$ (such that $\sum_{g=1}^{G}p_{g}=1$
and $p_{1}<...<p_{G}$) is a set of weights and $f_{N}\left( \cdot |\mu
_{g},\Omega _{g}\right) $ is the density of an $n$-variate Gaussian random
variable with mean $\mu _{g}$ and covariance matrix $\Omega _{g}$. In (\ref%
{mixture-normal}), we use the short-hand notation $\alpha =\left( \left(
p_{1},...,p_{G}\right) ^{\prime },\mu _{1}^{\prime },...,\mu _{G}^{\prime
},\left( vech\left( \Omega _{1}\right) \right) ^{\prime },...,\left(
vech\left( \Omega _{G}\right) \right) ^{\prime }\right) ^{\prime }$.

Finally, let now $\beta _{t}=\left( \beta _{1,t}^{\prime },...,\beta
_{n,t}^{\prime }\right) ^{\prime }$, $\omega =\left( b_{1}^{\prime
},...,b_{n}^{\prime }\right) ^{\prime }$, $A_{\beta }=\left\{ A_{\beta
,1},...,A_{\beta ,n}\right\} $, $\Sigma =\left\{ \Sigma _{1},...,\Sigma
_{n}\right\} $, and, for short,

\begin{equation*}
\theta =\left( vec\left( \alpha \right) ^{\prime },\omega ^{\prime
},vec\left( A_{\beta }\right) ^{\prime },vech\left( \Sigma ^{1/2}\right)
^{\prime },vec\left( \beta _{0}\right) ^{\prime }\right) ^{\prime }.
\end{equation*}%
Putting everything together, the resulting likelihood function (conditional
on the initial observations $\left\{ \mathbf{y}_{t}\right\} _{t=1}^{p}$) is
given by%
\begin{eqnarray}
L\left( \left\{ \mathbf{y}_{t}\right\} _{t=p+1}^{T}|\alpha ,\omega ,A_{\beta
},\Sigma ,\beta _{0},\left\{ \beta _{t}\right\} _{t=1}^{T}\right) 
&=&L\left( \left\{ \mathbf{y}_{t}\right\} _{t=p+1}^{T}|\theta ,\left\{ \beta
_{t}\right\} _{t=1}^{T}\right)   \label{tvp-likelihood} \\
&=&\prod\limits_{t=p+1}^{T}\left[ \prod\limits_{i=1}^{n}f_{i}\left(
y_{i,t}|z_{i,t};\beta _{i,t},A_{\beta ,i},\Sigma _{i},\beta
_{i,0},b_{i}\right) \right] c\left( \mathbf{y}_{t}|\alpha \right) ,  \notag
\end{eqnarray}%
where we have now emphasized the dependence of the marginal densities on $%
A_{\beta ,i}$, $\Sigma _{i}$ and $\beta _{i,0}$. 
It follows that 
\begin{equation}
\ln L\left( \left\{ \mathbf{y}_{t}\right\} _{t=p+1}^{T}|\theta ,\left\{
\beta _{t}\right\} _{t=1}^{T}\right) =\sum_{t=p+1}^{T}\ln c\left( \mathbf{y}%
_{t}|\alpha \right) +\sum_{i=1}^{n}\sum_{t=p+1}^{T}\ln f_{i}\left(
y_{i,t}|z_{i,t};\beta _{i,t},A_{\beta ,i},\Sigma _{i},\beta
_{i,0},b_{i}\right) ,  \label{log-likelihood}
\end{equation}%
which indicates that maximisation with respect to each unit $i$ can be
carried out separately, like maximisation with respect to $\alpha $.

\subsection{Dimension reduction and estimation\label{estimation}}

Equation (\ref{log-likelihood}) indicates that the likelihood can be
factored into $n+1$ independent problems. We choose the prior%
\begin{equation}
p\left( \theta \right) =p\left( \alpha \right) \prod\limits_{i=1}^{n}p\left(
\omega _{i,0}\right) p\left( A_{\beta ,i}\right) p\left( \Sigma _{i}\right)
p\left( \beta _{i,0}\right) ,  \label{tvp-prior}
\end{equation}%
where: $p\left( \alpha \right) $ and $p\left( \omega _{i,0}\right) $ are
flat priors (in the latter, coefficients are restricted to be non-negative); 
$p\left( \Sigma _{i}\right) \propto \left\vert \Sigma _{i}\right\vert
^{-\left( n+1\right) /2}$ as a standard non-informative prior; finally, $p\left( A_{\beta ,i}\right) $ and $p\left( \beta
_{i,0}\right) $ are Gaussian priors and we discuss them in details in Section \ref{prior-sensitivity}. Thus, by construction, $p\left( \theta
\right) $ can also be factorised into $n+1$ independent problems. 

Hence, the posterior is given by%
\begin{equation}
p\left( \theta ,\left\{ \beta _{t}\right\} _{t=1}^{T}|\left\{ \mathbf{y}%
_{t}\right\} _{t=1}^{T}\right) \propto L\left( \left\{ \mathbf{y}%
_{t}\right\} _{t=p+1}^{T}|\theta ,\left\{ \beta _{t}\right\}
_{t=1}^{T}\right) p\left( \theta \right) .  \label{tvp-posterior}
\end{equation}%
Note that, based on (\ref{tvp-likelihood}) and (\ref{tvp-prior}), the
posterior again factorises into separate posteriors for each unit-specific
set of parameter. This entails, as discussed in the introduction, that the
estimation of the TVP-VAR with possible heteroskedasticity\ can be
decomposed into $n+1$ estimation problems that can be carried out in
parallel, independently of each other. We point out that this result holds
as long as the prior on $\alpha $, $p\left( \alpha \right) $, is independent
of the other parameters; conversely, the prior on the other parameters can
have a hierarchical structure, so that (\ref{tvp-prior}) might alternatively be
written as%
\begin{equation*}
p\left( \theta \right) =p\left( \alpha \right) \prod\limits_{i=1}^{n}p\left(
A_{\beta ,i},\Sigma _{i},\beta _{i,0},b_{i}|\alpha \right) .
\end{equation*}%
Then, by standard arguments, (\ref{tvp-posterior}) would become%
\begin{align*}
p\left( \alpha |\left\{ \mathbf{y}_{t}\right\} _{t=1}^{T}\right) =&
\prod\limits_{t=p+1}^{T}c\left( \mathbf{y}_{t}|\alpha \right) p\left( \alpha
\right) , \\
p\left( A_{\beta ,i},\Sigma _{i},\beta _{i,0},b_{i},\left\{ \beta
_{t}\right\} _{t=1}^{T}|\left\{ \mathbf{y}_{t}\right\} _{t=1}^{T}\right) =&
\prod\limits_{t=p+1}^{T}f_{i}\left( \mathbf{y}_{t}|z_{t};A_{\beta ,i},\Sigma
_{i},\beta _{i,0},b_{i},\beta _{i,t}\right) p\left( A_{\beta ,i},\Sigma
_{i},\beta _{i,0},b_{i}|\alpha \right) p\left( \alpha \right) .
\end{align*}

Prior to discussing estimation, some considerations on the potential for
dimensionality reduction are in order. Despite the presence of the copula
term, the number of parameters in $\theta $ is still proportional to $n^{2}$%
, which does not fully resolve the challenge represented by dimensionality
in a large VAR. More specifically, from (\ref{mixture-normal}), it is
apparent that, when estimating $\mu _{g}$, the number of parameters to be
estimated is $Gn$; conversely, the matrices $\Omega _{g}$ contain each $%
\frac{n\left( n+1\right) }{2}$ elements and this is where the dimensionality
issue arises from. In order to attenuate this problem, in Section \ref{step2}
we consider two ways of restricting the $\Omega _{g}$s, which both reduce
the number of free parameters in the copula to being proportional to $n$ as
opposed to $n^{2}$.

\subsubsection{Univariate regressions estimation - $\beta_{i,t}$ and $b_{i}$\label{step1}}

Each equation (\ref{univ}) and (\ref{tvp-univ-2}) is a regression (or, if specification (\ref{univ}) is indeed chosen, an autoregression) with time varying parameters and stochastic volatility. Thus, we use the approach by \citet{kim1998} to estimate $\beta_{i,t}$ and $b_{i}$ (and the other hyperparameters).\\
More precisely, note that (\ref{garch}) entails that
\begin{equation}
\ln u_{i,t}^{2}=h_{i,t}+\ln u_{i,t}^{*2}.
\end{equation}

Thus, conditional on $\beta_{i,t}$s, we have
\begin{equation}
\ln\left(y_{i,t}-\sum_{j=1}^{p}\beta_{i,t}y_{i,t-j}\right)^{2}=h_{i,t}+\ln u_{i,t}^{*2}.
\end{equation}

The model is linear in $h_{i,t}$. It is well known (see \citealp{kim1998}) that using a Quasi-Maximum Likelihood estimator under the assumption that $\ln u_{i,t}^{*2}$ is normal
results in poor small-sample properties; thus, we follow the approach suggested by \citet{kim1998}. In particular, we approximate the
distribution of $\ln u_{i,t}^{*2}$ using a mixture of normals with
seven components. Thence,
for each $i$, $h_{i,t}$s is sampled at once using the Kalman filter. In turn, conditionally on $h_{i,t}$, the model for $y_{i,t}$ has a linear
state space representation in terms of $\beta_{i,t}$s. Therefore,
for each $i$, we draw the entire vector $\beta_{i,t}$ at once,
using again the Kalman filter.\footnote{We point out that an alternative approach is to use the Gibbs sampler
to draw from the conditional posterior distribution $p\left(\beta_{i,t}|\{h_{i,\tau},\tau\neq t\},\{h_{i,t}\},\{\mathbf{y}\}_{t=1}^{T}\right) $
but this approach, although simpler, results in slower convergence
and higher autocorrelation in MCMC draws.}

\subsubsection{Copula density estimation - $\protect\alpha $\label{step2}}

As is typical with copula models, we first obtain an estimate of the
univariate densities $f_{i}\left( y_{i,t}|z_{i,t};A_{\beta ,i},\Sigma
_{i},\beta _{i,0},b_{i},\beta _{i,t}\right) $. We then obtain the
probability integral transforms, $v_{i,t}$, and use these as data to
estimate $\alpha $.\footnote{%
This procedure can be viewed as \textquotedblleft
two-stage\textquotedblright Bayesian, as opposed to a \textquotedblleft
full-information\textquotedblright Bayesian estimator (see also %
\citealp{creal}). Whilst this could, in principle, be carried out by
modifying the MCMC algorithm, it adds to the computational complexity of the
estimation; further, we have tried to use it in some of our empirical
applications, but results were - if anything - marginally worse than with
the proposed two-step procedure which we study here.}

The dimensionality issue can be further addressed by imposing some
restrictions on $\alpha $. We discuss two possible approaches (denoted as 
\textit{S1} and \textit{S2}), where the priors employed are flat.

\bigskip

\textit{Dimension reduction: strategy S1}

\bigskip

The first dimension reduction strategy is based on a recursive model for the 
$\Omega _{g}$s:%
\begin{equation}
\Omega _{g}=h_{g}\Omega _{g-1}+V_{g},\text{ \ \ }2\leq g\leq G,  \label{s1}
\end{equation}%
having initialised (\ref{s1}) by leaving $\Omega _{1}$ unrestricted (and
thus setting $V_{1}=0$). In (\ref{s1}), $h_{g}$ is a scalar to be estimated,
and the idiosyncratic shock $V_{g}$ is restricted to be $V_{g}=diag\left\{
v_{g,1},...,v_{g,n}\right\} $.

Consequently, the number of parameters associated to the copula is $\left( G-1\right)
\left( n+1\right) $, which therefore grows linearly, as
opposed to quadratically, with $n$.

\bigskip

\textit{Dimension reduction: strategy S2}

\bigskip

Another possible dimension reduction approach is intimately related to
Principal Components (we refer to \citealp{humphreys} for a full treatment,
which we briefly summarize here), and to the Bayesian compression literature
(\citealp{dunson}). We again leave $\Omega _{1}$ unrestricted, and model the 
$\Omega _{g}$s, for $2\leq g\leq G$, as%
\begin{equation}
\Omega _{g}=Q_{0,g}Q_{0,g}^{\prime }+D_{g}.  \label{s2}
\end{equation}%
In (\ref{s2}), $D_{g}=diag\left\{ d_{g,1},...,d_{g,n}\right\} $ and $Q_{0,g}$
is an $n\times k$ matrix. \newline
We make no attempt to estimate $Q_{0,g}$. Instead, we randomly generate the
elements of $Q_{0,g}$, say $\left\{ Q_{0,g}\right\} _{i,j}$, $1\leq i\leq n$%
, $1\leq j\leq k$, as independent of each other with $\left\{
Q_{0,g}\right\} _{i,j}\sim N\left( 0,q_{g}^{2}\right) $ for a total of $%
1,000,000$ iterations, choosing the specification which maximises the log
marginal likelihood. Thus, the only parameters that need to be estimated are 
$q_{g}$ and $\left\{ d_{g,1},...,d_{g,n}\right\} $. Under these
restrictions, the number of parameters is $\left( G-1\right) \left( n+1\right) $ - i.e., the same as in \textit{S1}. 

\subsubsection{Sampling from the posterior: the MALA\ algorithm\label{mala}}

Sampling from (\ref{tvp-posterior}) can be done along similar lines as in
the case of a fixed coefficient VARs, but with the complications arising
from $\beta _{t}$ being time-varying. We use the Metropolis Adjusted
Langevin (MALA) algorithm by \citet{roberts1998} (see also \citealp{girolami}), which is likely to be more
efficient than an ordinary Random Walk Metropolis algorithm in light of the
large dimensionality of $\theta $.

In order to illustrate the algorithm, we begin by defining the matrix%
\begin{equation}
G\left( \widetilde{\theta }\right) =\left. -\frac{\partial ^{2}}{\partial
\theta \partial \theta ^{\prime }}\ln L\left( \left\{ \mathbf{y}_{t}\right\}
_{t=p+1}^{T}|\theta \right) \right\vert _{\theta =\widetilde{\theta }}
\label{hessian}
\end{equation}%
computed at a generic value $\widetilde{\theta }$. The likelihood $L\left(
\left\{ \mathbf{y}_{t}\right\} _{t=p+1}^{T}|\theta \right) $ is
differentiable up to any order, within the whole parameter space, due to the
normality assumption; thus, by the Schwarz Lemma, $G\left( \widetilde{\theta 
}\right) $ is symmetric for any $\widetilde{\theta }$ within the parameter
space.

Based on the definitions above, the resampling scheme is as follows:

\bigskip

\begin{description}
\item[\textit{GC-Step 1}] Initialise by drawing $\theta _{0}$ from $p\left(
\theta \right) $, and set $k=0$.

\item[\textit{GC-Step 2}] Randomly generate $\widetilde{\theta }$ from the
proposal density%
\begin{equation}
q\left( \widetilde{\theta }|\theta _{k}\right) \sim N\left[ m\left( \theta
_{k}\right) ,\lambda ^{2}I_{d}\right] .  \label{tvp-proposal}
\end{equation}

\item[\textit{GC-Step 3}] Compute the Metropolis acceptance probability%
\begin{equation}
A\left( \widetilde{\theta },\theta _{k}\right) =\min \left\{ 1,\frac{p\left( 
\widetilde{\theta }|\left\{ \mathbf{y}_{t}\right\} _{t=1}^{T}\right) }{%
p\left( \theta _{k}|\left\{ \mathbf{y}_{t}\right\} _{t=1}^{T}\right) }\frac{%
q\left( \theta _{k}|\widetilde{\theta }\right) }{q\left( \widetilde{\theta }%
|\theta _{k}\right) }\right\}   \label{acceptance}
\end{equation}

\item[\textit{GC-Step 4}] Draw $u$ from a uniform distribution in $\left[ 0,1%
\right] $, defining the acceptance rule%
\begin{equation*}
\begin{array}{c}
\text{if }u\leq A\left( \widetilde{\theta },\theta _{k}\right)
\Longrightarrow \theta _{k+1}=\widetilde{\theta } \\ 
\text{if }u>A\left( \widetilde{\theta },\theta _{k}\right) \Longrightarrow
\theta _{k+1}=\theta _{k}%
\end{array}%
\end{equation*}

\item[\textit{GC-Step 5}] Set $k=k+1$ and return to Step 2.
\end{description}

\bigskip

We now discuss the proposal density. In (\ref{tvp-proposal}), the scale
parameter $\lambda $ is discussed later on, and the mean $m\left( \theta
_{k}\right) $ is given by%
\begin{equation}
m\left( \theta _{k}\right) =\theta _{k}+\frac{1}{2}\lambda ^{2}\nabla \ln
p\left( \theta _{k}|\left\{ \mathbf{y}_{t}\right\} _{t=1}^{T}\right) ,
\label{tvp-mean}
\end{equation}%
where \textquotedblleft $\nabla $\textquotedblright\ refers to the gradient,
which is computed with respect to $\theta $ and then specialised in the
value $\theta _{k}$ (we use the same notation as \citealp{nemeth}). In (\ref%
{tvp-mean}), the main difficulty is the computation of 
\begin{equation}
\nabla \ln p\left( \theta _{k}|\left\{ \mathbf{y}_{t}\right\}
_{t=1}^{T}\right) =\nabla \ln L\left( \left\{ \mathbf{y}_{t}\right\}
_{t=1}^{T}|\theta _{k}\right) +\nabla \ln p\left( \theta _{k}\right) .
\label{grad-posterior}
\end{equation}%
Assuming - as is typical, see \citet{nemeth} - that $\nabla \ln p\left(
\theta _{k}\right) $ is known, this boils down to estimating $\nabla \ln
L\left( \left\{ \mathbf{y}_{t}\right\} _{t=1}^{T}|\theta _{k}\right) $. Note
that by Fisher's identity (\citealp{cappe}), it holds that%
\begin{equation}
\nabla \ln L\left( \left\{ \mathbf{y}_{t}\right\} _{t=1}^{T}|\theta
_{k}\right) =E_{\left\{ \beta _{t}\right\} _{t=1}^{T}}\left[ \nabla \ln
p\left( \left\{ \mathbf{y}_{t}\right\} _{t=1}^{T};\left\{ \beta _{t}\right\}
_{t=1}^{T}|\theta _{k}\right) \right] ,  \label{cappe}
\end{equation}%
where $E_{\left\{ \beta _{t}\right\} _{t=1}^{T}}$ denotes expectation taken
with respect to $p\left( \left\{ \beta _{t}\right\} _{t=1}^{T}|\left\{ 
\mathbf{y}_{t}\right\} _{t=1}^{T}\right) $, with 
\begin{equation}
\nabla \ln p\left( \left\{ \mathbf{y}_{t}\right\} _{t=1}^{T};\left\{ \beta
_{t}\right\} _{t=1}^{T}|\theta _{k}\right) =\sum_{t=1}^{T}\nabla \ln p\left( 
\mathbf{y}_{T}|\left\{ \mathbf{y}_{t}\right\} _{t=1}^{T-1};\left\{ \beta
_{t}\right\} _{t=1}^{T-1};\theta _{k}\right) +\nabla \ln p\left( \beta
_{T}|\left\{ \beta _{t}\right\} _{t=1}^{T-1};\theta _{k}\right) .
\label{grad-joint}
\end{equation}%
We carry out the estimation of $\nabla \ln L\left( \left\{ \mathbf{y}%
_{t}\right\} _{t=1}^{T}|\theta _{k}\right) $ using the Rao-Blackwellised
estimator proposed in \citet{nemeth}, as described below.

\bigskip

\begin{description}
\item[\textit{RB-Step 1}] Initialise by sampling the particles $\beta
_{1}^{\left( j\right) }$, $1\leq j\leq M$, from $p\left( \beta _{1}\right) $%
, and set%
\begin{equation*}
w_{1}^{\left( j\right) }=\frac{p\left( \mathbf{y}_{1}|\beta _{1}^{\left(
j\right) }\right) }{\sum_{j=1}^{M}p\left( \mathbf{y}_{1}|\beta _{1}^{\left(
j\right) }\right) },
\end{equation*}
computing also the estimate%
\begin{equation*}
\nabla \ln \widehat{L}\left( \mathbf{y}_{1}|\theta _{k}\right) =\nabla \ln
p\left( \mathbf{y}_{1}|\beta _{1}^{\left( j\right) };\theta _{k}\right)
+\nabla \ln p\left( \beta _{1}\right) .
\end{equation*}

\item[\textit{RB-Step 2}] For $t=2,...,T$, assume a set of weights $\left\{
\xi _{t}^{\left( j\right) }\right\} _{j=1}^{M}$ and a proposal density $%
q\left( \beta _{t}|\beta _{t-1}^{\left( j\right) };\mathbf{y}_{t};\theta
_{k}\right) $, and

\begin{description}
\item[\textit{(i)}] sample a set of indices $\left\{ k_{j}\right\}
_{j=1}^{M} $ from $1\leq j\leq M$, with probabilities $\left\{ \xi
_{t}^{\left( j\right) }\right\} _{j=1}^{M}$;

\item[\textit{(ii)}] define the updated weights%
\begin{equation*}
w_{t}^{\left( j\right) }=\frac{\widetilde{w}_{t}^{\left( j\right) }}{%
\sum_{j=1}^{M}\widetilde{w}_{t}^{\left( j\right) }},
\end{equation*}%
where%
\begin{equation*}
\widetilde{w}_{t}^{\left( j\right) }=\frac{\widetilde{w}_{t-1}^{\left(
k_{j}\right) }p\left( \mathbf{y}_{t}|\beta _{t}^{\left( j\right) };\theta
_{k}\right) p\left( \beta _{t}^{\left( j\right) }|\beta _{t-1}^{\left(
k_{j}\right) };\theta _{k}\right) }{\xi _{t}^{\left( k_{j}\right) }q\left(
\beta _{t}^{\left( j\right) }|\beta _{t-1}^{\left( k_{j}\right) };\mathbf{y}%
_{t};\theta _{k}\right) };
\end{equation*}

\item[\textit{(iii)}] compute%
\begin{equation}
m_{t}^{\left( j\right) }=\zeta m_{t-1}^{\left( k_{j}\right) }+\left( 1-\zeta
\right) \sum_{j=1}^{M}w_{t-1}^{\left( j\right) }m_{t-1}^{\left( j\right)
}+\ln p\left( \mathbf{y}_{t}|\beta _{t}^{\left( j\right) };\theta
_{k}\right) +\nabla \ln p\left( \beta _{t}^{\left( j\right) }|\beta
_{t-1}^{\left( k_{j}\right) };\theta _{k}\right) .  \label{nemeth-1}
\end{equation}
\end{description}

\item[\textit{RB-Step 3}] Compute%
\begin{equation*}
\nabla \ln \widehat{L}\left( \left\{ \mathbf{y}_{s}\right\}
_{s=1}^{t}|\theta _{k}\right) =\sum_{j=1}^{M}w_{t}^{\left( j\right)
}m_{t}^{\left( j\right) }.
\end{equation*}
\end{description}

\bigskip

The output of the algorithm is the estimate $\nabla \ln \widehat{L}\left(
\left\{ \mathbf{y}_{t}\right\} _{t=1}^{T}|\theta _{k}\right) $, which can
then be plugged in (\ref{grad-posterior}). As indicated by \citet{nemeth},
the algorithm also readily affords the computation of other important
quantities such as the predictive likelihood, etc.

\section{Empirical application\label{empirics}}

In this section, we illustrate our methodology by applying it to the estimation of a (large) TVP-VAR with heteroskedasticity. We use the same data as \citet{koop13},
namely $n=25$ US macroeconomic variables (see Table \ref%
{tab:TableKoop}) running from 1959:Q1 to 2010:Q2.
The focus of our exercise is the prediction of three series: inflation, GDP
and interest rate. Given that all series are transformed into first
differences in order to ensure stationarity, our model predicts the
percentage change in inflation (the second log difference of CPI), GDP
growth (the log difference of real GDP) and the change in the interest rate
(the difference of the Fed funds rate). To ensure a fair comparison with %
\citet{koop13}, we have also demeaned all variables and then standardised
them (we use the standard deviation calculated from 1959Q1 through 1969Q4).
The forecasting horizon is 1970:Q1 till 2010:Q2.

Results are in Tables \ref{tab:Table3}-\ref{tab:Table5}, where we report the
relative Mean Squared Forecast Errors (MSFE) when using the various VAR
specifications to predict GDP, inflation and interest rate (respectively).
The numbers in the tables are the MSFE relative to the TVP-VAR-DMA model in %
\citet{koop13}, which is therefore our baseline model.

\begin{table}[h!]
\centering
\begin{threeparttable}
{\tiny

\caption{Relative MSFE at various
horizons $h$ - predictions for
GDP.}
\label{tab:Table3}\centering
\par
\begin{tabular}{lllllllll}  \\
[0.2cm] 
\hline 
GDP & \multicolumn{8}{c}{Forecast horizon} \\[0.25cm] 
&
$h=1$ & $h=2$ & $h=3$ & $h=4$ & $h=5$ & $h=6$ & $h=7$ & $h=8$ \\ \hline \\
[0.25cm] 
TVP-VAR$^{\left( a\right) }$, $\lambda =0.99$, $\beta _{T+h}\sim
RW$ & 1.17
& 1.17 & 1.10 & 1.12 & 1.12 & 1.10 & 1.13 & 1.12 \\ [0.25cm]

TVP-VAR$^{\left( a\right) }$, $\lambda =0.99$, $\kappa =0.96$, $\alpha
=0.99$
& 1.04 & 1.05 & 1.01 & 1.02 & 1.01 & 1.00 & 1.02 & 1.01 \\ [0.25cm]

VAR, Heteroskedastic$^{\left( a\right) }$ & 1.10 & 1.10 & 1.03 & 1.04 &
1.05
& 1.01 & 1.06 & 1.04 \\ [0.25cm] 
VAR, Homoskedastic$^{\left(
a\right) }$ & 1.13 & 1.03 & 1.03 & 1.05 & 1.08 & 
1.06 & 1.10 & 1.08 \\
[0.25cm]
\hline \\ [0.25cm] 
TVP-VAR, this paper & 1.12 & 1.14 & 1.15 &
1.17 & 1.20 & 1.20 & 1.22 & 1.25
\\ [0.25cm] 
TVP-VAR, using \textit{S1} &
0.90 & 0.92 & 0.94 & 0.94 & 0.96 & 0.98 & 1.00
& 1.02 \\ [0.25cm]

TVP-VAR, using \textit{S2} & 0.89 & 0.91 & 0.92 & 0.92 & 0.90 & 0.94 &
0.92
& 0.97 \\ [0.25cm] 
VAR - Heteroskedastic, this paper ($G=1$) & 1.01
& 0.99 & 0.99 & 0.97 & 0.98
& 0.98 & 1.01 & 1.03 \\ [0.25cm] 
VAR -
Heteroskedastic, GMCM & 0.91 & 0.93 & 0.95 & 0.99 & 1.02 & 1.04 & 
1.07 &
1.09 \\ [0.25cm] 
VAR - Heteroskedastic, Bayes compression & 0.84 & 0.87 &
0.87 & 0.90 & 0.90
& 0.92 & 0.92 & 0.93 \\ [0.25cm] 
VAR - Homoskedastic,
this paper ($G=1$) & 0.94 & 0.96 & 0.99 & 1.01 & 1.05 & 
1.07 & 1.09 & 1.14
\\ [0.25cm] 
VAR - Homoskedastic, GMCM & 0.89 & 0.91 & 0.91 & 0.92 & 0.94 &
0.95 & 0.97
& 0.97 \\ [0.25cm] 
VAR - Homoskedastic, Bayes compression &
0.90 & 0.91 & 0.93 & 0.93 & 0.95 & 
0.98 & 1.09 & 1.14 \\
 \hline \\
[0.25cm] 
\end{tabular}
\begin{tablenotes}
      \tiny
    \item In each
column, $h$ denotes the horizon for which the prediction has been computed.

    \item The first panel of the table contains the results for several
models proposed in \citet{koop13}; specifically, the superscript
\textquotedblleft $^{\left( a\right) }$\textquotedblright refers to the
models considered in Table 1 in  \citet{koop13}. In the first row,
\textquotedblleft $RW$\textquotedblright denotes a random walk law of motion
for the time-varying parameters; the parameters in the second row are
defined in \citet{koop13}.
    \item In the second panel of the table, we
use the models proposed in this paper. In particular, in the model denoted
as \textquotedblleft GMCM\textquotedblright, we use the mixed Gaussian
copula model defined in (\ref{mixture-normal}); $G$ has been selected equal
to 4 based on the values of the marginal likelihood. In the row above, we
have used $G=1$, with no model selection. In the row denoted as
\textquotedblleft Bayes compression\textquotedblright, we have used the
methodology proposed by \citet{dunson}, averaging across $10,000$ sets of
weights, derived from marginal likelihoods converted into posterior
probabilities.    
\end{tablenotes}
}  
\end{threeparttable}
\end{table}
\bigskip

\begin{table}[h!]
\centering
\begin{threeparttable}
{\tiny

\caption{Relative MSFE at various
horizons $h$ - predictions for
inflation.}
\label{tab:Table4}\centering
\par
\begin{tabular}{lllllllll} 
\\ [0.2cm] 
\hline 
Inflation & \multicolumn{8}{c}{Forecast horizon} \\
[0.25cm] 
& $h=1$ & $h=2$ & $h=3$ & $h=4$ & $h=5$ & $h=6$ & $h=7$ & $h=8$
\\ \hline \\[0.25cm] 
TVP-VAR$^{\left( a\right) }$, $\lambda =0.99$, $\beta
_{T+h}\sim RW$ & 1.03
& 1.02 & 1.00 & 1.01 & 1.00 & 1.00 & 1.00 & 1.02 \\
[0.25cm] 
TVP-VAR$^{\left( a\right) }$, $\lambda =0.99$, $\kappa =0.96$,
$\alpha =0.99$
& 1.03 & 1.02 & 1.03 & 1.04 & 1.00 & 1.00 & 1.02 & 1.00 \\
[0.25cm] 
VAR, Heteroskedastic$^{\left( a\right) }$ & 1.03 & 1.02 & 1.01 &
1.02 & 1.01
& 1.00 & 1.01 & 1.02 \\ [0.25cm] 
VAR, Homoskedastic$^{\left(
a\right) }$ & 1.04 & 1.06 & 1.03 & 1.02 & 1.00 & 
1.03 & 1.01 & 1.01 \\
[0.25cm] \hline \\[0.25cm] 
TVP-VAR, this paper & 0.85 & 0.92 & 0.97 & 1.00
& 1.00 & 1.02 & 1.02 & 1.04
\\ [0.25cm] 
TVP-VAR, using \textit{S1} & 0.95
& 0.96 & 0.97 & 0.98 & 1.00 & 1.00 & 1.01
& 1.03 \\ [0.25cm] 
TVP-VAR,
using \textit{S2} & 0.93 & 0.95 & 0.95 & 0.97 & 0.97 & 0.98 & 1.00
& 1.00
\\ [0.25cm] 
VAR - Heteroskedastic, this paper ($G=1$) & 0.81 & 0.83 & 0.83
& 0.85 & 0.87
& 0.87 & 0.89 & 0.92 \\ [0.25cm] 
VAR - Heteroskedastic,
GMCM & 0.74 & 0.74 & 0.75 & 0.75 & 0.77 & 0.77 & 
0.82 & 0.84 \\ [0.25cm]

VAR - Heteroskedastic, Bayes compression & 0.97 & 0.99 & 1.00 & 1.00 &
1.02
& 1.04 & 1.07 & 1.09 \\ [0.25cm] 
VAR - Homoskedastic, this paper
($G=1$) & 1.01 & 1.03 & 1.03 & 1.05 & 1.07 & 
1.07 & 1.09 & 1.13 \\
[0.25cm] 
VAR - Homoskedastic, GMCM & 0.82 & 0.82 & 0.80 & 0.82 & 0.94 &
0.97 & 1.00
& 1.03 \\ [0.25cm] 
VAR - Homoskedastic, Bayes compression &
1.05 & 1.05 & 1.07 & 1.09 & 1.10 & 
1.10 & 1.12 & 1.15 \\ \hline \\[0.25cm]

\end{tabular}
\begin{tablenotes}
      \tiny
      \item The models
considered in the table are the same as in Table
\ref{tab:Table3}.
\end{tablenotes}
}  
\end{threeparttable}
\end{table}
\bigskip

\begin{table}[h!]
\centering
\begin{threeparttable}
{\tiny

\caption{Relative MSFE at various
horizons $h$ - predictions for interest
rates.}
\label{tab:Table5}\centering
\par
\begin{tabular}{lllllllll}  \\
[0.2cm] 
\hline 
Interest Rate & \multicolumn{8}{c}{Forecast horizon} \\
[0.25cm] 
& $h=1$ & $h=2$ & $h=3$ & $h=4$ & $h=5$ & $h=6$ & $h=7$ & $h=8$
\\ \hline \\[0.25cm] 
TVP-VAR$^{\left( a\right) }$, $\lambda =0.99$, $\beta
_{T+h}\sim RW$ & 1.11
& 1.03 & 1.02 & 1.02 & 1.02 & 1.02 & 1.01 & 0.99 \\
[0.25cm] 
TVP-VAR$^{\left( a\right) }$, $\lambda =0.99$, $\kappa =0.96$,
$\alpha =0.99$
& 1.10 & 1.09 & 1.05 & 1.08 & 1.02 & 1.01 & 1.03 & 1.02 \\
[0.25cm] 
VAR, Heteroskedastic$^{\left( a\right) }$ & 1.10 & 1.01 & 1.01 &
1.02 & 1.01
& 1.01 & 1.03 & 1.03 \\ [0.25cm] 
VAR, Homoskedastic$^{\left(
a\right) }$ & 1.11 & 1.07 & 1.11 & 1.11 & 1.03 & 
1.03 & 1.09 & 1.08 \\
[0.25cm]  \hline \\[0.25cm] 
TVP-VAR, this paper & 0.93 & 0.95 & 0.95 &
0.97 & 1.00 & 1.00 & 1.01 & 1.03
\\ [0.25cm] 
TVP-VAR, using \textit{S1} &
0.90 & 0.92 & 0.93 & 0.94 & 0.95 & 0.99 & 1.02
& 1.03 \\ [0.25cm]

TVP-VAR, using \textit{S2} & 0.91 & 0.93 & 0.93 & 0.95 & 0.97 & 0.97 &
0.97
& 1.03 \\ [0.25cm] 
VAR - Heteroskedastic, this paper ($G=1$) & 0.90
& 0.92 & 0.92 & 0.94 & 0.94
& 0.96 & 0.98 & 0.99 \\ [0.25cm] 
VAR -
Heteroskedastic, GMCM & 0.83 & 0.85 & 0.86 & 0.86 & 0.86 & 0.88 & 
0.91 &
0.93 \\ [0.25cm] 
VAR - Heteroskedastic, Bayes compression & 0.80 & 0.80 &
0.81 & 0.83 & 0.83
& 0.85 & 0.87 & 0.90 \\ [0.25cm] 
VAR - Homoskedastic,
this paper ($G=1$) & 0.94 & 0.95 & 0.97 & 0.97 & 1.03 & 
1.05 & 1.07 & 1.08
\\ [0.25cm] 
VAR - Homoskedastic, GMCM & 0.87 & 0.91 & 0.91 & 0.93 & 0.96 &
0.99 & 1.01
& 1.04 \\ [0.25cm] 
VAR - Homoskedastic, Bayes compression &
0.97 & 0.97 & 1.00 & 1.02 & 1.02 & 
1.03 & 1.05 & 1.07 \\ \hline \\[0.25cm]

\end{tabular}
\begin{tablenotes}
      \tiny
     
           \item
The models considered in the table are the same as in Table
\ref{tab:Table3}.
\end{tablenotes}
}  
\end{threeparttable}
\end{table}
\bigskip

Broadly speaking, results show that our methodology affords good forecasting
ability especially for shorter horizons; a notable exception is the poor
performance of the TVP-VAR for GDP, although using strategies \textit{S1}
and \textit{S2} yields a marked improvement. Indeed, there is no clearly
superior model, although the results seem to make a case for heteroskedastic
VARs (nonetheless, homoskedastic VARs with GMCM show very good results). In
general, using GMCM (and determining the $G$) works better than restricting $%
G$ to 1 (as could be expected). Similarly, reducing the dimensionality of
the copula model with either strategy \textit{S1} or \textit{S2} generally
improves forecasting ability. Although Bayesian compression works well, it
does not seem to yield a uniformly superior predictive performance than the
univariate models proposed in equation (\ref{univ}). As a final point, a
distinctive feature of \citet{koop13} is that the authors propose to use
\textquotedblleft forgetting factors\textquotedblright (a procedure not dissimilar to an exponentially weighted
moving average); thus, they avoid estimating the covariance matrix of the
VAR and the covariance matrix of the time-varying coefficients. In our case,
we are dealing with univariate models, and therefore we do not have to
estimate covariance matrices.

\subsection{Prior sensitivity \label{prior-sensitivity}}

We have carried out a further exercise to explore the sensitivity of our
methodology to the choice of the (main) priors on $A_{\beta,i}$ and $\beta_{i,0}$. We point out that - in this contribution - the main focus is not so much the choice of the prior but the copula-based dimensionality reduction. Indeed, we propose flat priors in general, although of course some parameters undergo nonlinear transformations which invalidates this argument (see the classical reference by \citealp{jeffreys} for an early treatment of the issue). Hence the importance of at least validating the choice of our priors through sensitivity analysis.

We begin by describing the benchmark prior. For each element in the vector $\left(vec\left( A_{\beta,i}\right)', \beta_{i,0}' \right)'$, we have chosen the prior $N(\overline{b},\overline{s}_{b}^{2})$, independent across elements. As far as the copula
functions are concerned, recall (\ref{mixture-normal}). We have used both
dimension reduction strategies \textit{S1} and \textit{S2}, with:%
\begin{equation*}
p_{g}=\frac{e^{-r_{g}^{2}}}{\sum_{g=1}^{G}e^{-r_{g}^{2}}},
\end{equation*}%
where 
\begin{equation}
p\left( r_{g}\right) =N(\overline{r},\overline{s}_{r}^{2}),  \label{p-rg}
\end{equation}%
and 
\begin{equation}
p\left( \mu _{g}\right) =N(\overline{m}_{g},\overline{s}_{m}^{2}),
\label{p-mg}
\end{equation}%
again independent for $1\leq g\leq G$. Finally, in (\ref{s2}), we have used $%
\Omega _{g}=C_{g}C_{g}^{\prime }$, with : 
\begin{equation}
p\left( c_{g}\right) =N(\overline{c},\overline{s}_{c}^{2}),  \label{p-cg}
\end{equation}%
for $2\leq g\leq G$, where we have defined $c_{g}=vech\left( C_{g}\right) $.
We have set the priors parameters as follows: 
\begin{equation}
\begin{array}{c}
\overline{b}=0,\overline{s}_{b}^{2}=10, \\ 
\overline{r}=0,\overline{s}_{r}^{2}=100, \\ 
\overline{m}_{g}=0,\overline{s}_{m}^{2}=10, \\ 
\overline{c}=0,\overline{s}_{c}^{2}=100.%
\end{array}
\label{prior parameters}
\end{equation}

\bigskip

In our analysis, we have used $1,000$ different priors by sampling randomly
from (\ref{p-rg})-(\ref{p-cg}), given the parameters defined in (\ref%
{prior parameters}). For each prior, we have used MCMC sampling, employing $%
10,000$ iterations starting from the posterior moments delivered by the
benchmark prior. Note that we have not examined sensitivity with respect to
other priors, which are anyway rather diffuse.

\bigskip

In order to compare our results against the TVP-VAR-DMA model in \citet{koop13}, we have computed the relative MSFEs as above.
The sampling distributions of these relative MSFEs are reported in Figures %
\ref{Figure5}-\ref{Figure6}. As can be seen, strategy \textit{S2} seems to
deliver the best results, both in terms of the mode of the sampling
distribution, and the dispersion around it.

\section{Conclusions\label{conclusions}}

Our paper has developed an alternative methodology for the direct
estimation of large TVP-VARs with possible heteroskedasticity. The original multivariate model
is decomposed into $n$ simpler models, whose interactions are modelled
separately through a copula. We use the GMCM copula, whose good performance
in our context is in line with the conclusions of other papers (see e.g. %
\citealp{geweke2007} and \citealp{villani}). In principle, however, it would
be possible to use also other copula specifications; given that considering this approaches goes beyond the scope of our paper (and the GMCM copula did not pose any particular runtime issues), this is an area for future research.

Our empirical
applications (see also the estimation of VARMAs in Appendix \ref{varma}) show that our approach is computationally more convenient
than directly estimating multivariate models. In addition, our results also
show excellent goodness of fit and predictive ability. We note that, when
reducing the original multivariate model into $n$ separate models, it is not necessary to impose a pure AR(1) structure in which each series is predicted
using solely its own lags. Indeed, we also consider a different model
reduction strategy based on Bayesian compression. However, we found that
even univariate, simple AR(1) models afford good forecasting ability. These
considerations support the conclusion that the use of copulas,
particularly in high dimension, is advantageous in that the copula manages
to capture features of the data that the original, standard multivariate
models are likely to miss.
Thus, our contribution may also be viewed as a complement to the recent advances in the Bayesian
analysis of large VARs, such as the ones developed in \citet{banbura} and %
\citet{glp}, where - instead of using copulas - new, more sophisticated
priors are proposed as a way to deal with large VARs.

We point out that our applications mainly focus on \textquotedblleft reduced
form\textquotedblright\ examples, as can be seen by the emphasis on
forecasting ability. We conjecture however that, in light of its excellent
performance, our technique could also be employed
in the context of more structural applications. This issue is currently under
examination by the authors.

\bibliographystyle{chicago}
\bibliography{ITT_biblio}

\begin{thebibliography}{}

\bibitem[\protect\citeauthoryear{Ba{\'n}bura, Giannone, and
  Reichlin}{Ba{\'n}bura et~al.}{2010}]{banbura}
Ba{\'n}bura, M., D.~Giannone, and L.~Reichlin (2010).
\newblock Large {B}ayesian vector auto regressions.
\newblock {\em Journal of Applied Econometrics\/}~{\em 25\/}(1), 71--92.

\bibitem[\protect\citeauthoryear{Bassetti, Casarin, and Leisen}{Bassetti
  et~al.}{2014}]{bassetti2014beta}
Bassetti, F., R.~Casarin, and F.~Leisen (2014).
\newblock Beta-product dependent {P}itman--{Y}or processes for {B}ayesian
  inference.
\newblock {\em Journal of Econometrics\/}~{\em 180\/}(1), 49--72.

\bibitem[\protect\citeauthoryear{Canova}{Canova}{1993}]{canova}
Canova, F. (1993).
\newblock Modelling and forecasting exchange rates with a {B}ayesian
  time-varying coefficient model.
\newblock {\em Journal of Economic Dynamics and Control\/}~{\em 17\/}(1-2),
  233--261.

\bibitem[\protect\citeauthoryear{Capp{\'e}, Moulines, and Ryd{\'e}n}{Capp{\'e}
  et~al.}{2009}]{cappe}
Capp{\'e}, O., E.~Moulines, and T.~Ryd{\'e}n (2009).
\newblock Inference in hidden {M}arkov models.
\newblock In {\em Proceedings of EUSFLAT conference}, pp.\  14--16.

\bibitem[\protect\citeauthoryear{Carriero, Clark, and Marcellino}{Carriero
  et~al.}{2015}]{carriero2015}
Carriero, A., T.~E. Clark, and M.~Marcellino (2015).
\newblock Bayesian {VAR}s: specification choices and forecast accuracy.
\newblock {\em Journal of Applied Econometrics\/}~{\em 30\/}(1), 46--73.

\bibitem[\protect\citeauthoryear{Carriero, Clark, and Marcellino}{Carriero
  et~al.}{2016}]{carriero2012}
Carriero, A., T.~E. Clark, and M.~Marcellino (2016).
\newblock Common drifting volatility in large {B}ayesian {VAR}s.
\newblock {\em Journal of Business \& Economic Statistics\/}~{\em 34\/}(3),
  375--390.

\bibitem[\protect\citeauthoryear{Carriero, Clark, and Marcellino}{Carriero
  et~al.}{2019}]{carriero2016large}
Carriero, A., T.~E. Clark, and M.~Marcellino (2019).
\newblock Large bayesian vector autoregressions with stochastic volatility and
  non-conjugate priors.
\newblock {\em Journal of Econometrics\/}~{\em 212}, 137--154.

\bibitem[\protect\citeauthoryear{Chan, Eisenstat, et~al.}{Chan
  et~al.}{2013}]{chan2013}
Chan, J.~C., E.~Eisenstat, et~al. (2013).
\newblock Gibbs samplers for {VARMA} and its extensions.
\newblock Technical report, Australian National University, College of Business
  and Economics.

\bibitem[\protect\citeauthoryear{Chan, Eisenstat, and Koop}{Chan
  et~al.}{2016}]{chan2015}
Chan, J.~C., E.~Eisenstat, and G.~Koop (2016).
\newblock Large bayesian {VARMA}s.
\newblock {\em Journal of Econometrics\/}~{\em 192\/}(2), 374--390.

\bibitem[\protect\citeauthoryear{Chen and Huang}{Chen and
  Huang}{2007}]{chen2007nonparametric}
Chen, S.~X. and T.-M. Huang (2007).
\newblock Nonparametric estimation of copula functions for dependence
  modelling.
\newblock {\em Canadian Journal of Statistics\/}~{\em 35\/}(2), 265--282.

\bibitem[\protect\citeauthoryear{Chib and Greenberg}{Chib and
  Greenberg}{1996}]{chib1996}
Chib, S. and E.~Greenberg (1996).
\newblock Markov chain monte carlo simulation methods in econometrics.
\newblock {\em Econometric Theory\/}~{\em 12\/}(3), 409--431.

\bibitem[\protect\citeauthoryear{Clark}{Clark}{2011}]{clark2011}
Clark, T.~E. (2011).
\newblock Real-time density forecasts from bayesian vector autoregressions with
  stochastic volatility.
\newblock {\em Journal of Business \& Economic Statistics\/}~{\em 29\/}(3),
  327--341.

\bibitem[\protect\citeauthoryear{Clark and Ravazzolo}{Clark and
  Ravazzolo}{2015}]{clark2015}
Clark, T.~E. and F.~Ravazzolo (2015).
\newblock Macroeconomic forecasting performance under alternative
  specifications of time-varying volatility.
\newblock {\em Journal of Applied Econometrics\/}~{\em 30\/}(4), 551--575.

\bibitem[\protect\citeauthoryear{Cogley and Sargent}{Cogley and
  Sargent}{2001}]{cogley}
Cogley, T. and T.~J. Sargent (2001).
\newblock Evolving post-{W}orld {W}ar {II US} inflation dynamics.
\newblock {\em NBER macroeconomics annual\/}~{\em 16}, 331--373.

\bibitem[\protect\citeauthoryear{Creal and Tsay}{Creal and Tsay}{2015}]{creal}
Creal, D.~D. and R.~S. Tsay (2015).
\newblock High dimensional dynamic stochastic copula models.
\newblock {\em Journal of Econometrics\/}~{\em 189\/}(2), 335--345.

\bibitem[\protect\citeauthoryear{Di~Lucca, Guglielmi, M{\"u}ller, and
  Quintana}{Di~Lucca et~al.}{2013}]{di2013simple}
Di~Lucca, M.~A., A.~Guglielmi, P.~M{\"u}ller, and F.~A. Quintana (2013).
\newblock A simple class of {B}ayesian nonparametric autoregression models.
\newblock {\em Bayesian Analysis (Online)\/}~{\em 8\/}(1), 63.

\bibitem[\protect\citeauthoryear{Doan, Litterman, and Sims}{Doan
  et~al.}{1984}]{doan}
Doan, T., R.~Litterman, and C.~Sims (1984).
\newblock Forecasting and conditional projection using realistic prior
  distributions.
\newblock {\em Econometric Reviews\/}~{\em 3\/}(1), 1--100.

\bibitem[\protect\citeauthoryear{Geweke and Amisano}{Geweke and
  Amisano}{2014}]{geweke}
Geweke, J. and G.~Amisano (2014).
\newblock Analysis of variance for {B}ayesian inference.
\newblock {\em Econometric Reviews\/}~{\em 33\/}(1-4), 270--288.

\bibitem[\protect\citeauthoryear{Geweke and Keane}{Geweke and
  Keane}{2007}]{geweke2007}
Geweke, J. and M.~Keane (2007).
\newblock Smoothly mixing regressions.
\newblock {\em Journal of Econometrics\/}~{\em 138\/}(1), 252--290.

\bibitem[\protect\citeauthoryear{Giannone, Lenza, and Primiceri}{Giannone
  et~al.}{2015}]{glp}
Giannone, D., M.~Lenza, and G.~E. Primiceri (2015).
\newblock Prior selection for vector autoregressions.
\newblock {\em Review of Economics and Statistics\/}~{\em 97\/}(2), 436--451.

\bibitem[\protect\citeauthoryear{Giannone and Reichlin}{Giannone and
  Reichlin}{2006}]{giannone2006}
Giannone, D. and L.~Reichlin (2006).
\newblock Does information help recovering structural shocks from past
  observations?
\newblock {\em Journal of the European Economic Association\/}~{\em 4\/}(2-3),
  455--465.

\bibitem[\protect\citeauthoryear{Girolami and Calderhead}{Girolami and
  Calderhead}{2011}]{girolami}
Girolami, M. and B.~Calderhead (2011).
\newblock Riemann manifold {L}angevin and {H}amiltonian {M}onte {C}arlo
  methods.
\newblock {\em Journal of the Royal Statistical Society: Series B (Statistical
  Methodology)\/}~{\em 73\/}(2), 123--214.

\bibitem[\protect\citeauthoryear{Gouri{\'e}roux, Monfort, and
  Renne}{Gouri{\'e}roux et~al.}{2019}]{gourierouxmonfort}
Gouri{\'e}roux, C., A.~Monfort, and J.-P. Renne (2019).
\newblock Identification and estimation in non-fundamental structural {VARMA}
  models.
\newblock {\em The Review of Economic Studies\/}, to appear.

\bibitem[\protect\citeauthoryear{Gruber and Czado}{Gruber and
  Czado}{2015}]{gruber2015}
Gruber, L. and C.~Czado (2015).
\newblock Sequential {B}ayesian model selection of regular vine copulas.
\newblock {\em Bayesian Analysis\/}~{\em 10\/}(4), 937--963.

\bibitem[\protect\citeauthoryear{Gruber and Czado}{Gruber and
  Czado}{2018}]{gruber2018}
Gruber, L.~F. and C.~Czado (2018).
\newblock Bayesian model selection of regular vine copulas.
\newblock {\em Bayesian Analysis\/}~{\em 13\/}(4), 1107--1131.

\bibitem[\protect\citeauthoryear{Guhaniyogi and Dunson}{Guhaniyogi and
  Dunson}{2015}]{dunson}
Guhaniyogi, R. and D.~B. Dunson (2015).
\newblock Bayesian compressed regression.
\newblock {\em Journal of the American Statistical Association\/}~{\em
  110\/}(512), 1500--1514.

\bibitem[\protect\citeauthoryear{Humphreys, Harris, Rodr{\'\i}guez-Higuero,
  Mubarak, Zhao, and Ojasalo}{Humphreys et~al.}{2015}]{humphreys}
Humphreys, D.~A., P.~M. Harris, M.~Rodr{\'\i}guez-Higuero, F.~A. Mubarak,
  D.~Zhao, and K.~Ojasalo (2015).
\newblock Principal component compression method for covariance matrices used
  for uncertainty propagation.
\newblock {\em IEEE Transactions on Instrumentation and Measurement\/}~{\em
  64\/}(2), 356--365.

\bibitem[\protect\citeauthoryear{Ibragimov}{Ibragimov}{2005}]{ibragimov}
Ibragimov, R. (2005).
\newblock Copula-based dependence characteriztions and modeling for time
  series.
\newblock {\em Harvard Institute of Economic Research Discussion
  Paper\/}~(2094).

\bibitem[\protect\citeauthoryear{Jeffreys}{Jeffreys}{1998}]{jeffreys}
Jeffreys, H. (1998).
\newblock {\em The theory of probability}.
\newblock OUP Oxford.

\bibitem[\protect\citeauthoryear{Kapetanios, Marcellino, and
  Venditti}{Kapetanios et~al.}{2019}]{venditti}
Kapetanios, G., M.~Marcellino, and F.~Venditti (2019).
\newblock Large time-varying parameter {VAR}s: A non-parametric approach.
\newblock {\em Journal of Applied Econometrics\/}, to appear.

\bibitem[\protect\citeauthoryear{Kim, Shephard, and Chib}{Kim
  et~al.}{1998}]{kim1998}
Kim, S., N.~Shephard, and S.~Chib (1998).
\newblock Stochastic volatility: likelihood inference and comparison with
  {ARCH} models.
\newblock {\em The Review of Economic Studies\/}~{\em 65\/}(3), 361--393.

\bibitem[\protect\citeauthoryear{Koop and Korobilis}{Koop and
  Korobilis}{2013}]{koop13}
Koop, G. and D.~Korobilis (2013).
\newblock Large time-varying parameter {VAR}s.
\newblock {\em Journal of Econometrics\/}~{\em 177\/}(2), 185--198.

\bibitem[\protect\citeauthoryear{Koop, Korobilis, and Pettenuzzo}{Koop
  et~al.}{2019}]{koop19}
Koop, G., D.~Korobilis, and D.~Pettenuzzo (2019).
\newblock Bayesian compressed vector autoregressions.
\newblock {\em Journal of Econometrics\/}~{\em 210\/}(1), 135--154.

\bibitem[\protect\citeauthoryear{Korobilis and Pettenuzzo}{Korobilis and
  Pettenuzzo}{2019}]{korobilis2019adaptive}
Korobilis, D. and D.~Pettenuzzo (2019).
\newblock Adaptive hierarchical priors for high-dimensional vector
  autoregressions.
\newblock {\em Journal of Econometrics\/}~{\em 212}, 241--271.

\bibitem[\protect\citeauthoryear{Litterman}{Litterman}{1986}]{litterman1986}
Litterman, R.~B. (1986).
\newblock Forecasting with {B}ayesian vector autoregressions: five years of
  experience.
\newblock {\em Journal of Business \& Economic Statistics\/}~{\em 4\/}(1),
  25--38.

\bibitem[\protect\citeauthoryear{L{\"u}tkepohl}{L{\"u}tkepohl}{2005}]{lutkepohl91}
L{\"u}tkepohl, H. (2005).
\newblock {\em New introduction to multiple time series analysis}.
\newblock Springer Science \& Business Media.

\bibitem[\protect\citeauthoryear{L{\"u}tkepohl and Poskitt}{L{\"u}tkepohl and
  Poskitt}{1996}]{poskitt}
L{\"u}tkepohl, H. and D.~S. Poskitt (1996).
\newblock Specification of echelon-form {VARMA} models.
\newblock {\em Journal of Business \& Economic Statistics\/}~{\em 14\/}(1),
  69--79.

\bibitem[\protect\citeauthoryear{Nelsen}{Nelsen}{2007}]{nelsen}
Nelsen, R.~B. (2007).
\newblock {\em An introduction to copulas}.
\newblock Springer Science \& Business Media.

\bibitem[\protect\citeauthoryear{Nemeth, Sherlock, and Fearnhead}{Nemeth
  et~al.}{2016}]{nemeth}
Nemeth, C., C.~Sherlock, and P.~Fearnhead (2016).
\newblock Particle {M}etropolis-adjusted {L}angevin algorithms.
\newblock {\em Biometrika\/}~{\em 103\/}(3), 701--717.

\bibitem[\protect\citeauthoryear{Nieto-Barajas, M{\"u}ller, Ji, Lu, and
  Mills}{Nieto-Barajas et~al.}{2012}]{nieto2012time}
Nieto-Barajas, L.~E., P.~M{\"u}ller, Y.~Ji, Y.~Lu, and G.~B. Mills (2012).
\newblock A time-series {DDP} for functional proteomics profiles.
\newblock {\em Biometrics\/}~{\em 68\/}(3), 859--868.

\bibitem[\protect\citeauthoryear{Nieto-Barajas and Quintana}{Nieto-Barajas and
  Quintana}{2016}]{nieto2016bayesian}
Nieto-Barajas, L.~E. and F.~A. Quintana (2016).
\newblock A {B}ayesian non-parametric dynamic {AR} model for multiple time
  series analysis.
\newblock {\em Journal of Time Series Analysis\/}~{\em 37\/}(5), 675--689.

\bibitem[\protect\citeauthoryear{Primiceri}{Primiceri}{2005}]{primiceri}
Primiceri, G.~E. (2005).
\newblock Time varying structural vector autoregressions and monetary policy.
\newblock {\em The Review of Economic Studies\/}~{\em 72\/}(3), 821--852.

\bibitem[\protect\citeauthoryear{Roberts and Rosenthal}{Roberts and
  Rosenthal}{1998}]{roberts1998}
Roberts, G.~O. and J.~S. Rosenthal (1998).
\newblock Optimal scaling of discrete approximations to {L}angevin diffusions.
\newblock {\em Journal of the Royal Statistical Society: Series B (Statistical
  Methodology)\/}~{\em 60\/}(1), 255--268.

\bibitem[\protect\citeauthoryear{Rodriguez and ter Horst}{Rodriguez and ter
  Horst}{2008}]{rodriguez2008bayesian}
Rodriguez, A. and E.~ter Horst (2008).
\newblock Bayesian dynamic density estimation.
\newblock {\em Bayesian Analysis\/}~{\em 3\/}(2), 339--365.

\bibitem[\protect\citeauthoryear{Scaillet and Fermanian}{Scaillet and
  Fermanian}{2003}]{scaillet2002}
Scaillet, O. and J.-D. Fermanian (2003).
\newblock Nonparametric estimation of copulas for time series.
\newblock {\em Journal of Risk\/}~(5), 25--54.

\bibitem[\protect\citeauthoryear{Sims}{Sims}{1980}]{sims80}
Sims, C.~A. (1980).
\newblock Macroeconomics and reality.
\newblock {\em Econometrica\/}~{\em 48}, 1--48.

\bibitem[\protect\citeauthoryear{Sims}{Sims}{1993}]{sims93}
Sims, C.~A. (1993).
\newblock A nine-variable probabilistic macroeconomic forecasting model.
\newblock In {\em Business Cycles, Indicators and Forecasting}, pp.\  179--212.
  University of Chicago Press.

\bibitem[\protect\citeauthoryear{Sklar}{Sklar}{1996}]{sklar2}
Sklar, A. (1996).
\newblock Random variables, distribution functions, and copulas: a personal
  look backward and forward.
\newblock {\em Lecture notes-monograph series\/}, 1--14.

\bibitem[\protect\citeauthoryear{Sklar}{Sklar}{1959}]{sklar1}
Sklar, M. (1959).
\newblock Fonctions de repartition an dimensions et leurs marges.
\newblock {\em Publ. Inst. Statist. Univ. Paris\/}~{\em 8}, 229--231.

\bibitem[\protect\citeauthoryear{Stock and Watson}{Stock and
  Watson}{1996}]{stock96}
Stock, J.~H. and M.~W. Watson (1996).
\newblock Evidence on structural instability in macroeconomic time series
  relations.
\newblock {\em Journal of Business \& Economic Statistics\/}~{\em 14\/}(1),
  11--30.

\bibitem[\protect\citeauthoryear{Taddy}{Taddy}{2010}]{taddy2010autoregressive}
Taddy, M.~A. (2010).
\newblock Autoregressive mixture models for dynamic spatial {P}oisson
  processes: Application to tracking intensity of violent crime.
\newblock {\em Journal of the American Statistical Association\/}~{\em
  105\/}(492), 1403--1417.

\bibitem[\protect\citeauthoryear{Tewari, Giering, and Raghunathan}{Tewari
  et~al.}{2011}]{tewari}
Tewari, A., M.~J. Giering, and A.~Raghunathan (2011).
\newblock Parametric characterization of multimodal distributions with
  non-gaussian modes.
\newblock In {\em 2011 IEEE 11th International Conference on Data Mining
  Workshops}, pp.\  286--292. IEEE.

\bibitem[\protect\citeauthoryear{Tsionas, Izzeldin, and Trapani}{Tsionas
  et~al.}{2019}]{itt2019}
Tsionas, M., M.~Izzeldin, and L.~Trapani (2019).
\newblock Copula-based {B}ayesian estimation of large multivariate stochastic
  volatility models.
\newblock Technical report.

\bibitem[\protect\citeauthoryear{Uhlig}{Uhlig}{1997}]{uhlig}
Uhlig, H. (1997).
\newblock Bayesian vector autoregressions with stochastic volatility.
\newblock {\em Econometrica\/}~{\em 65}, 59--74.

\bibitem[\protect\citeauthoryear{Villani, Kohn, and Giordani}{Villani
  et~al.}{2009}]{villani}
Villani, M., R.~Kohn, and P.~Giordani (2009).
\newblock Regression density estimation using smooth adaptive {G}aussian
  mixtures.
\newblock {\em Journal of Econometrics\/}~{\em 153\/}(2), 155--173.

\bibitem[\protect\citeauthoryear{Watson}{Watson}{1994}]{watson94}
Watson, M.~W. (1994).
\newblock Vector autoregressions and cointegration.
\newblock {\em Handbook of Econometrics\/}~{\em 4}, 2843--2915.

\end{thebibliography}
\newpage

\clearpage
\newpage
\appendix
\setcounter{section}{0} \setcounter{subsection}{-1} \setcounter{equation}{0} %
\setcounter{lemma}{0} \setcounter{table}{0} \setcounter{figure}{0} %
\renewcommand{\thelemma}{A.\arabic{lemma}} \renewcommand{\theequation}{A.%
\arabic{equation}} \renewcommand{\thetable}{A.\arabic{table}} %
\renewcommand{\thefigure}{A.\arabic{figure}}

\section{Tables and figures\label{figures}}

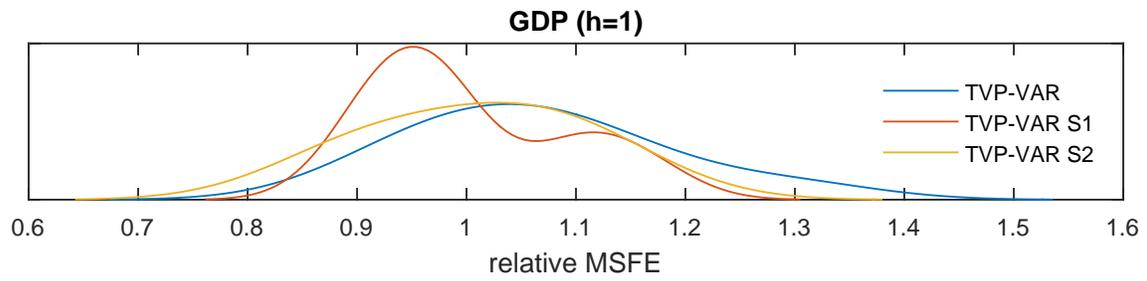
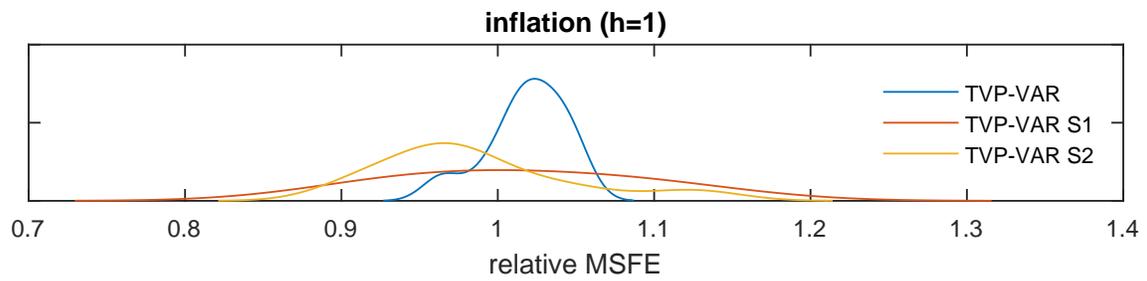
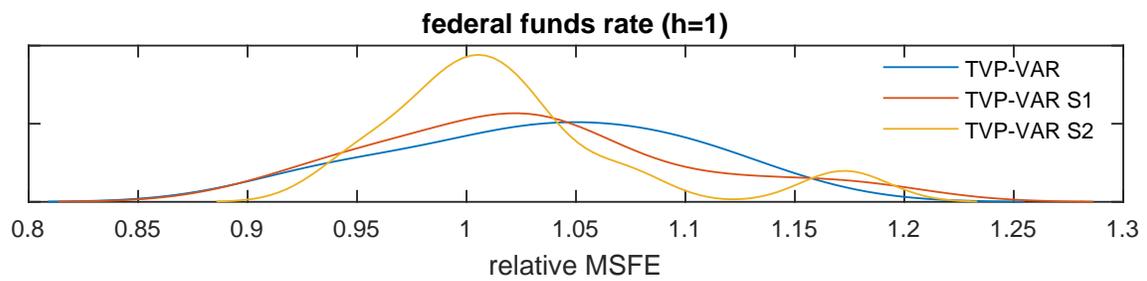
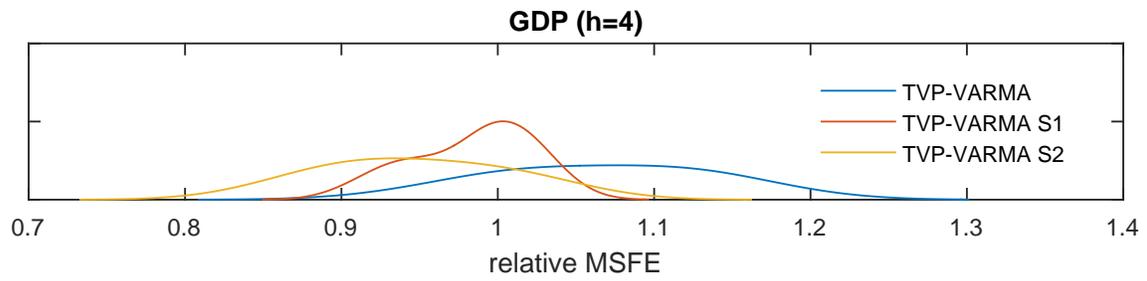
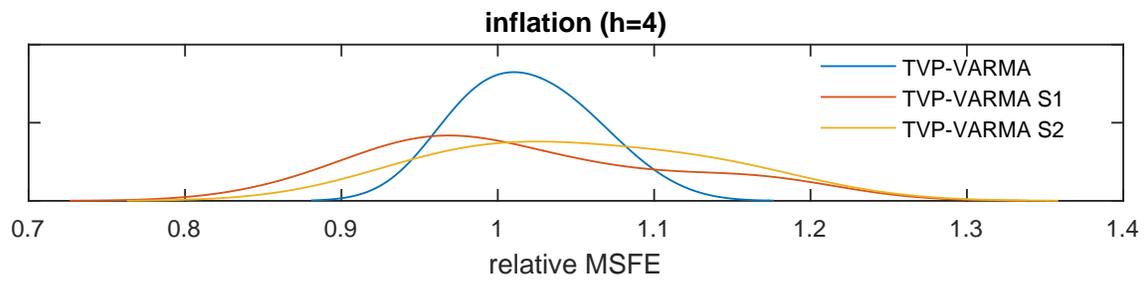
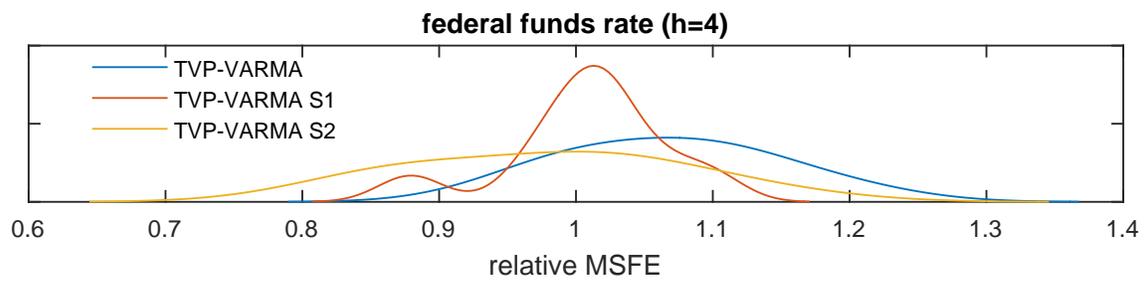
\begin{table}[h!]
\caption{List of variables employed in \citet{koop13}}
\label{tab:TableKoop}\centering
\par
\begin{tabular}{lllll}
&  &  &  &  \\[0.2cm] \hline
&  &  &  &  \\[0.25cm] 
GDP & Industrial production & US/UK exchange rate &  &  \\[0.25cm] 
CPI & Capacity utilisation & Real personal consumption expenditures &  &  \\%
[0.25cm] 
Fed Funds rate & Unemployment rate & Total nonfarm payroll &  &  \\[0.25cm] 
NAPM CPI & Housing starts & ISM Manifacturing (PMI\ composite) &  &  \\%
[0.25cm] 
Borrowing from Fed & Producer price index & ISM Manifacturing (New orders) & 
&  \\[0.25cm] 
S\&P500 & Average hourly earnings & Output per hour &  &  \\[0.25cm] 
M2 money stock & M1 money stock &  &  &  \\[0.25cm] 
Personal income & Spot oil price &  &  &  \\[0.25cm] 
Real GDPI & 10-year T-bill &  &  &  \\[0.25cm] \hline
&  &  &  &  \\[0.25cm] 
&  &  &  & 
\end{tabular}%
\end{table}

\begin{figure}[h!]
\caption{Sampling distributions of the relative MSFEs (forecasting horizon: $%
h=1$ periods) - the benchmark is the TVP-VAR-DMA model in \citet{koop13}}
\label{Figure5}\centering
\includegraphics[width=\textwidth]{prior1}
\end{figure}

\begin{figure}[h!]
\caption{Sampling distributions of the relative MSFEs (forecasting horizon: $%
h=4$ periods) - the benchmark is the TVP-VAR-DMA model in \citet{koop13}}
\label{Figure6}\centering
\includegraphics[width=\textwidth]{prior2}
\end{figure}

\clearpage
\newpage

\setcounter{section}{1} \setcounter{subsection}{0} \setcounter{equation}{0} %
\setcounter{lemma}{0} \setcounter{table}{0} \setcounter{figure}{0} %
\renewcommand{\thelemma}{B.\arabic{lemma}} \renewcommand{\theequation}{B.%
\arabic{equation}} \renewcommand{\thetable}{B.\arabic{table}} %
\renewcommand{\thefigure}{B.\arabic{figure}}

\section{Application to a VARMA using US data\label{varma}}

In order to further illustrate the flexibility and the performance of our approach, we consider the estimation and the predictive ability of a VARMA model, applied to US macro data. Our exercise is based on \citet{chan2015}, who make a compelling case for the use of VARMAs, in light of their superior predictive ability (see also the theory in \citealp{poskitt}). Yet, VARMAs, as well as suffering from well-known identification issues (see e.g. the recent contribution by \citealp{gourierouxmonfort}) are liable to overparameterisation, and therefore dimensionality, in this context, is a very important issue. \newline
In this application, we do not consider time variation: the purpose of our analysis is only to show the computational advantages of our procedure. 
We follow \citet{chan2015}, using the same dataset. The data are quarterly
US macroeconomic variables, ranging from 1959:Q1 to 2013:Q4. All data are
first-differenced to obtain stationarity, as is customarily recommended in
this type of analysis (see \citealp{carriero2015}) - see Table \ref{tab:TableChan} for a
list of variables employed. In order to ensure a meaningful comparison with %
\citet{chan2015}, we consider three models of increasing dimension, with $%
n=3,7$ and $12$. In particular, for
the model with $n=3$ variables, we have used: Real GDP, CPI (All Items) and
Effective Federal Funds Rate. For the case $n=7$, the variables are the ones
in the previous model, plus: Average Hourly Earnings: Manufacturing, M2
Money Stock, Spot Oil Price (WTI), and S\&P 500 Index. Finally, for the
model with $n=12$, the additional variables are Real Personal Consumption,
Housing Starts (total), Real GPDI, ISM PMI Composite Index and All Employees
(Total nonfarm).\footnote{%
A complete description of the dataset is available from the authors.}

\begin{table}[h!]
\caption{List of variables employed in \citet{chan2015}}
\label{tab:TableChan}\centering
\par
\begin{tabular}{lllll}
&  &  &  &  \\[0.2cm] \hline
&  &  &  &  \\[0.25cm] 
GDP & GDP & GDP &  &  \\[0.25cm] 
CPI (All Items) & CPI (All Items) & CPI (All Items) &  &  \\[0.25cm] 
Effective Fed Fund Rate & Effective Fed Fund Rate & Effective Fed Fund Rate
&  &  \\[0.25cm] 
& Average Hourly Earnings (Manifacturing) & Average Hourly Earnings
(Manifacturing) &  &  \\[0.25cm] 
& M2 & M2 &  &  \\[0.25cm] 
& Spot Oil Price (WTI) & Spot Oil Price (WTI) &  &  \\[0.25cm] 
& S\&P500 Index & S\&P500 Index &  &  \\[0.25cm] 
&  & Real Personal Consumption &  &  \\[0.25cm] 
&  & Housing Starts (total) &  &  \\[0.25cm] 
&  & Real GDPI &  &  \\[0.25cm] 
&  & ISM PMI Composite Index &  &  \\[0.25cm] 
&  & All Employees (total nonfarm) &  &  \\[0.25cm] \hline
&  &  &  &  \\[0.25cm] 
&  &  &  & 
\end{tabular}%
\end{table}

From a methodological point of view, given the posterior $p\left( \theta
\right) $, we employ the same resampling scheme as suggested in %
\citet{girolami}; the algorithm is essentially the same as the one reported in
Section \ref{mala}. The only difference is the proposal density
employed in \textit{GC-Step 2}, which in the case of a fixed parameter
VARMA\ is defined as 
\begin{equation}
q\left( \widetilde{\theta }|\theta _{k}\right) \sim N\left( \theta
_{k},\varepsilon ^{2}G^{-1}\left( \theta _{k}\right) \right) ,
\label{proposal-girolami}
\end{equation}%
where $\varepsilon $ is chosen so as to pre-determine, roughly, the
acceptance ratio in \textit{GC-Step 4} of the algorithm, setting it to around $25-30
$. We do this using $600,000$ replications, with a burn-in period of $100,000
$ replications.

\bigskip 

In Table \ref{tab:Table1}, we compare models using the sum of the log
predictive likelihood as a model selection criterion based on forecasting
accuracy (see also an insightful contribution by \citealp{geweke}).\footnote{%
Details and formulas (also for other indicators) are available upon request.} As can be noted, our methodology yields results which, broadly speaking, are
as good as the ones in \citet{chan2015}; a distinctive advantage is that our
procedure is simpler and quicker to implement (CPU times are always below 1'
using mainframe). Similarly to \citet{chan2015}, we note that VARMA models
seem to offer better predictive power; yet, remarkably, our VAR($4$) based
on using the dimension reduction strategy denoted as \textit{S2} is at least
as good (in fact, marginally better) than both our VARMA($4,4 $), and the
one in \citet{chan2015}. Interestingly, it can be noted that model averaging
yields an even better result. This can be viewed as an indication that none
of the models under consideration is correctly specified, which makes the
case for model averaging.

\begin{table}[h]
\centering
\begin{threeparttable}
{\tiny
\caption{Sum of log predictive
likelihoods for various
specifications}
\label{tab:Table1}
\par
\begin{tabular}{lllll} 
\\ [0.2cm] 
\hline \\ [0.25cm]
& $n=3$ & $n=7$ & $n=12$ & $W$ \\ 
&  &  &
&  \\ 
\textbf{\citet{chan2015}} &  &  &  &  \\[0.25cm]

$VARMA(4,4)^{\left( a\right) }$ & $-182.5$ & $-401.9$ & $-492.3$ & 
\\[0.25cm] 
$VARMA(4,4)^{\left( b\right) }$ & $-188.1$ & $-406.0$ &
$-504.2$ &  \\[0.25cm] 
$VAR(4)$ & $-187.1$ & $-406.7$ & $-496.9$ & 
\\[0.25cm] 
&  &  &  &  \\ 
\hline \\ [0.25cm]
\textbf{This paper} &  & 
&  &  \\[0.25cm] 
$VAR(4)$ & $-187.1$ & $-406.8$ & $-496.9$ & $0.24$
\\[0.25cm] 
$VARMA(4,4)$ & $-182.5$ & $-401.9$ & $-492.1$ & $0.29$
\\[0.25cm] 
$VAR(4)$ \textit{with }$S1$ & $-187.0$ & $-406.7$ & $-496.7$ &
$0.22$ \\[0.25cm] 
$VAR(4)$ \textit{with }$S2$ & $-182.3$ & $-401.7$ &
$-492.0$ & $0.21$ \\[0.25cm] 
\textit{Model average} & $-179.4$ & $-401.1$
& $-490.0$ & $0.32$\\[0.25cm]
\hline 
\end{tabular} 

\begin{tablenotes}
  
\tiny
      \item The table contains the sums of the log predictive
likelihood for various specifications - in panel \textquotedblleft
{\citet{chan2015}}\textquotedblright, we consider various VARMA
specifications using the methodology proposed in \citet{chan2015} - the
superscripts  \textquotedblleft $^{\left( a\right) }$\textquotedblright and 
\textquotedblleft $^{\left( b\right) }$\textquotedblright refer to two
different prior specifications; in panel \textquotedblleft This
paper\textquotedblright, we have considered various specifications based on
our methodology.
      \item When using the two strategies \textit{S1} and
\textit{S2} described in Section \ref{step2}, $G$ has been selected by
maximising the integrated likelihood as a selection criterion; in all cases,
we this has led to the choice $G=3$.
      \item In the \textquotedblleft
Model average\textquotedblright row, we use a standard Bayesian model
averaging, based on weights computed from the posterior model probabilities
(details are available upon request); we have used $10,000$ sets of weights.

      \item The column denoted as $W$ contains the inefficiency factor of
our MCMC - see e.g. \citet{chib1996} for a definition. 
   
\end{tablenotes}
}
\end{threeparttable}
\end{table}

\bigskip

We have also run a complementary exercise, in which we estimate the three
models (with $n=3,7$ and $12$ variables), and then consider the predictive
likelihood for the variables that are common to the three specifications -
that is, Real GDP, CPI (All Items) and Effective Federal Funds Rate. The
results are presented in Table \ref{tab:Table2}, where it can be noted that
the forecasting ability (slightly) improves as $n$ increases, reinforcing
the case for large VARs. Especially when $n=12$ is considered, our approach
to the estimation of VARMA($4,4$) delivers the best predictive ability -
again, this result should be read in conjunction with the decidedly lower
CPU time of our approach.
\begin{table}[h!]
\centering
\begin{threeparttable}
{\tiny
\caption{Sum of log predictive
likelihoods - predictions of GDP, CPI and interest
rates}
\label{tab:Table2}\centering
\par
\begin{tabular}{llll}  \\
[0.2cm] 
\hline \\ [0.25cm]
& $n=3$ & $n=7$ & $n=12$ \\ 
&  &  &  \\

\textbf{\citet{chan2015}} &  &  &  \\[0.25cm] 
&  &  &  \\

$VARMA(4,4)^{\left( a\right) }$ & $-182.5$ & $-182.2$ & $-181.1$
\\[0.25cm] 
$VARMA(4,4)^{\left( b\right) }$ & $-188.1$ & $-185.4$ &
$-187.4$ \\[0.25cm] 
$VAR(4)$ & $-187.1$ & $-187.2$ & $-191.0$ \\[0.25cm]

&  &  &  \\ 
\hline \\ [0.25cm]
\textbf{This paper} &  &  &  \\[0.25cm]

$VAR(4)$ & $-183.1$ & $-184.2$ & $-189.0$ \\[0.25cm] 
$VARMA(4,4)$ &
$-180.5$ & $-180.2$ &
$-180.0$\\[0.25cm]
\hline
\end{tabular}
\begin{tablenotes}
      \tiny
  
\item The table contains the sum of the log predictive likelihoods based on
the predictive densities of the first three variables (Real GDP, CPI and
Interest rate). All other specifications are the same as in Table
\ref{tab:Table1}.
      \item Note that we do not report the weighted
average, since the posterior model probabilities favour only one model. 
  
\end{tablenotes}
}  
\end{threeparttable}

\end{table}
\bigskip

Finally, we point out that we have also carried out impulse response
analysis; whilst we do not report results (which are available upon
request), we found that impulse response functions behave in a very
similar way to those in Figures 1 and 2 in \citet{chan2015}.

\end{document}